\documentclass[showpacs,pre,aps,twocolumn,superscriptaddress]{revtex4-1}

\pdfoutput=1
\usepackage{amsmath,amsfonts,amssymb,bm,graphicx,hyperref,color,units}

\newcommand{\C}{\mathcal{C}}
\newcommand{\W}{\mathbb{W}}
\newcommand{\Ham}{\mathbb{H}}
\newcommand{\B}{\mathcal{B}}
\newcommand{\E}{\mathcal{E}}
\newcommand{\llangle}{\langle\!\langle}
\newcommand{\rrangle}{\rangle\!\rangle}
\newcommand{\bra}[1]{\langle #1|}
\newcommand{\ket}[1]{|#1\rangle}

\begin{document}

\title{Trajectory phase transitions and dynamical Lee-Yang zeros\\ of the Glauber-Ising chain}
\author{James M. Hickey}
\affiliation{School of Physics and Astronomy, University of Nottingham, Nottingham, NG7 2RD, United Kingdom}
\author{Christian Flindt}
\affiliation{D\'epartement de Physique Th\'eorique, Universit\'e de Gen\`eve, 1211 Gen\`eve, Switzerland}
\author{Juan P. Garrahan}
\affiliation{School of Physics and Astronomy, University of Nottingham, Nottingham, NG7 2RD, United Kingdom}

\date{\today}

\begin{abstract}
We examine the generating function of the time-integrated energy for the one-dimensional Glauber-Ising model. At long times, the generating function takes on a large-deviation form and the associated cumulant generating function has singularities corresponding to continuous trajectory (or ``space-time'') phase transitions between paramagnetic trajectories and ferromagnetically or anti-ferromagnetically ordered trajectories. In the thermodynamic limit, the singularities make up a whole curve of critical points in the complex plane of the counting field. We evaluate analytically the generating function by mapping the generator of the biased dynamics to a non-Hermitian Hamiltonian of an associated quantum spin chain. We relate the trajectory phase transitions to the high-order cumulants of the time-integrated energy which we use to extract the dynamical Lee-Yang zeros of the generating function. This approach offers the possibility to detect continuous trajectory phase transitions from the finite-time behaviour of measurable quantities.
\end{abstract}


%05.40.-a Fluctuation phenomena, random processes, noise, and Brownian motion
%64.70.P- Glass transitions
%72.70.+m Noise processes and phenomena

\maketitle

\section{Introduction}
\label{sec:Intro}

The dynamics of a complex many-body system may be much more intriguing than one would expect based on its equilibrium properties only~\cite{Barrat2004,*Hinrichsen2000}. This dynamical richness can be revealed by considering strictly dynamical observables that depend on the full time evolution of the system. The fluctuations of the dynamical observables capture temporal correlations in the dynamics, unlike static quantities which depend only on the state of the system at a given time~\cite{Chandler1987,*Sachdev2011}.  The complete statistical characterization of the dynamical observables encodes the dynamical properties of the system at hand.  This observation has led to the emergence of the field of full counting statistics (FCS) through parallel developments in quantum optics~\cite{Plenio1998,*gardiner2004}, electronic transport~\cite{Nazarov2003,*Esposito2009}, and classical stochastic processes~\cite{Gardiner1986,*VanKampen2007}. The typical observable is the total count of dynamical events such as the number of photons emitted from an atomic system~\cite{Cook1981,*Lenstra1982}, electrons having passed through a mesoscopic conductor~\cite{Levitov1993,*Levitov1996,*Pilgram2003,*Flindt2008}, or molecular motions in a glassy liquid~\cite{Hedges2009}, for which one aims to obtain the full statistical distribution.

To characterize and understand dynamical observables it is useful to pursue a statistical description using the language of equilibrium statistical mechanics~\cite{Ruelle2004}. Within this framework, one may think of dynamical trajectories in FCS as microstates in thermodynamics~\cite{Lecomte2007,Merolle2005}. The thermodynamic large-system-size limit then corresponds to the limit of long times in FCS, where the full statistical properties of the dynamical observables are captured by large-deviation (LD) functions~\cite{[{For reviews see }]Eckmann1985,*Gaspard2005,*Touchette2009} that play the role of dynamical free energies~\cite{Merolle2005,Garrahan2007}. An important consequence of this analogy is that the dynamical free energies, just as in the static case, may display singular changes in the dynamical fluctuations corresponding to trajectory phase transitions~\cite{Ates2012,Hickey2012,Li2011,Budini2011}. It has for instance been argued that a trajectory phase transition underlies the glass transition in liquid systems~\cite{Garrahan2007,Hedges2009,Pitard2011,Speck2012,Bodineau2012a}.

However, in contrast to equilibrium statistical mechanics, where a phase transition may be observed by tuning the physical field that drives it, for example a magnetic field in a magnetic system or the pressure in a fluid, the variable conjugate to the counted observable in FCS, the ``counting field'', is typically not directly related to any physically accessible parameters~\cite{Lecomte2007,Garrahan2007,Hedges2009,Pitard2011,Levkivskyi2009,Ivanov2010,Ivanov2013,Utsumi2013}. This makes it difficult, if not impossible, to observe trajectory phase transitions in experiment or even in simulations (without resorting to sophisticated sampling techniques~\cite{Garrahan2009}).  Furthermore, singularities in the dynamical free energy, corresponding to trajectory phase transitions, only appear in the limit of long times, i.~e.~at times that are much longer than the typical relaxation times, which often is beyond reach in practice.

Recently, two of us proposed a potential solution to these problems by establishing a general relation between trajectory phase transitions in stochastic many-body systems and the high-order cumulants of the corresponding dynamical observables~\cite{Flindt2013}. Specifically, we made use of a dynamic generalization of ideas from equilibrium statistical mechanics by Yang and Lee~\cite{Lee1952,*Yang1952,[{Other dynamical applications of Lee-Yang ideas have  for example been considered in }]Blythe2002,*Bena2005,*Wei2012} and connected the time evolution of the zeros of the moment generating function (MGF) for the dynamical observable of interest with the short-time behaviour of its high-order cumulants~\cite{Flindt2009,Flindt2010,Kambly2011}. As we showed, it is possible to infer the position of these dynamical Lee-Yang zeros from the high-order cumulants of the dynamical observable as they move towards the value of the counting field at which the trajectory phase transition occurs. The method was applied to two kinetically constrained models of glassy systems, the Frederickson-Andersen model~\cite{Fredrickson1984} and the East model~\cite{Jackle1991}, for which we showed that a first-order trajectory phase transition occurring at zero counting field can be inferred from numerical simulations of the short-time high-order cumulants of the activity; the number of configuration changes in the systems.

The purpose of the present work is to apply the proposed method to a stochastic many-body system which, unlike the models above, exhibits a continuous trajectory phase transition and furthermore has a whole curve of critical points in the complex plane of the counting field, rather than just a single transition point at zero counting field~\cite{Gorissen2009,Elmatad2010,Garrahan2011}. Concretely, we investigate trajectory phase transitions in the one-dimensional Glauber-Ising model \cite{Glauber1963} using the high-order cumulants of the time-integrated energy~\cite{Jack2010}. In contrast to our previous work, where the high-order cumulants were obtained from numerical simulations of the stochastic dynamics, the Glauber-Ising chain permits an analytical treatment as we shall see. This, in turn, allows us to test our method on a challenging problem and benchmark the results against exact solutions.

The Glauber-Ising chain exhibits a continuous trajectory phase transition from paramagnetic trajectories to ferromagnetically or anti-ferromagnetically ordered trajectories in the limits of long times and large system size~\cite{Jack2010,Loscar2011}. Here we demonstrate how the  trajectory phase diagram, with the counting field treated as a complex variable, can be extracted from the short-time evolution of the high-order cumulants resolved with respect to the individual modes of an associated non-Hermitian quantum Hamiltonian. Moreover, at low temperatures a few critical points dominate the high-order cumulants. This shows that one may use short-time cumulants to infer a continuous trajectory phase transition occurring at non-zero counting fields, even if the real physical dynamics takes place at zero counting field.

The paper is structured as follows: In Sec.~\ref{sec:Form} we outline the formalism used throughout the paper. In Sec.~\ref{sec:Method} we describe the method developed in Ref.~\cite{Flindt2013}. The Glauber-Ising model is discussed in Sec.~\ref{sec:Model} where we also calculate analytically the time-dependent MGF of the time-integrated energy. The corresponding trajectory phase diagram is introduced in Sec.~\ref{sec:dpg}. In Sec.~\ref{sec:kres} we proceed to present results for the high-order cumulants of the time-integrated energy and discuss how the full trajectory phase transition curve can be determined using mode-resolved cumulants. In Sec.~\ref{sec:MGFanaly} we perform a full analysis without using the mode-resolved cumulants and show that the dominating singularities of the system can still be extracted but only at low temperatures, where the fluctuations are dominated by the long-wavelength modes. In contrast, at high temperatures all modes are important, making it difficult to extract the transition points from the cumulants. Finally, in Sec.~\ref{sec:Conc} we present our conclusions and provide an outlook on extensions of our method and applications to other interesting systems for future work.

\section{$s$-ensemble Formalism}
\label{sec:Form}

We investigate trajectory phase transitions using a thermodynamic formalism known as the ``$s$-ensemble''. Within this framework, the stochastic trajectories of a complex many-body system are classified according to an associated time-extensive quantity $\B_t$ and its time-dependent probability distribution $P(\B_t)$~\cite{Garrahan2009}. Additionally, we introduce the time-intensive counting field $s$ conjugate to $\B_t$ by defining the MGF
\begin{equation}
\label{eq:MGF}
Z(s,t)=\sum_{\B_t}{e}^{-s\B_t}P(\B_t).
\end{equation}
The MGF yields the moments of $\B_t$ by differentiation with respect to the counting field at $s=0$:
\begin{equation}
\label{eq:moments}
\langle \B_t^{n}\rangle = (-1)^{n}\partial^{n}_{s}Z(s,t)|_{s\rightarrow0}.
\end{equation}
We define the cumulant generating function (CGF) as
\begin{equation}
\label{eq:CFG}
\Theta(s,t)=\log{Z(s,t)}
\end{equation}
with corresponding cumulants reading
\begin{equation}
\label{eq:cumu}
\llangle \B_t^{n}\rrangle = (-1)^{n}\partial^{n}_{s}\Theta(s,t)|_{s\rightarrow0}.
\end{equation}
In the limit of long times, the MGF obeys a LD principle and takes on the form \cite{[{For reviews see }]Eckmann1985,*Gaspard2005,*Touchette2009}
\begin{equation}
Z(s,t)\approx e^{t\theta(s)},\,\,\, t\rightarrow \infty,
\end{equation}
such that the CGF becomes
\begin{equation}
\label{eq:CGFlongtime}
\Theta(s,t)\approx t\theta(s),\,\,\, t\rightarrow \infty,
\end{equation}
where $\theta(s)$ is the LD function of the counting process.

Pursuing the analogy with equilibrium statistical mechanics~\cite{Lecomte2007}, we treat the counting variable $s$ as a thermodynamic field, and consider $\theta(s)$ as the corresponding dynamical free energy. Within this thermodynamic formalism, time plays the extensive role of volume for the counting process and the corresponding entropy density associated with the counting process may be obtained via a Legendre transformation. However, rather than tuning the equilibrium configuration of microstates as in equilibrium statistical mechanics, the counting field $s$ biases the trajectory ensemble from the actual one at $s = 0$ and may in some cases drive the system across a trajectory phase transition. Similar to equilibrium statistical mechanics, the analytic
properties of the dynamical free energy $\theta(s)$ encode information about the phase behaviour of ensembles of trajectories of the counting process~\cite{Garrahan2010,Garrahan2011,Genway2012,Hickey2012}.

The following discussion pertains to a large class of stochastic many-body problems. Here we focus for concreteness on systems that evolve according to a Master equation of the form
\begin{equation}
\label{eq:Meq}
\frac{d}{dt}P(\C,t) = -r(\C)P(\C,t)+\sum_{\C'}W(\C'\rightarrow \C)P(\C',t),
\end{equation}
where $P(\C,t)$ is the probability for the system to be in configuration $\C$ at time $t$. We denote the transition rate from configuration $\C$ to $\C'$ by $W(\C\rightarrow\C')$ and $r(\C)=\sum_{\C'}W(\C\rightarrow\C')$ is the total escape rate from configuration $\C$. Equation~(\ref{eq:Meq}) defines a system of linear differential equations for the $P(\C,t)$'s, which in a convenient matrix notation can be expressed as
\begin{equation}
\frac{d}{dt}\ket{P(t)}=\W\ket{P(t)}.
\end{equation}
The vector $\ket{P(t)}$ contains the probabilities $P(\C,t)$'s, while the matrix elements of $\W$ are
\begin{equation}
\label{eq:Meq2}
[\W]_{\{\C,\C'\}}= W(\C'\rightarrow\C)-r(\C)\delta_{\C,\C'}.
\end{equation}

In Ref.~\cite{Flindt2013} we classified the stochastic trajectories according to their dynamical activity, i.~e.\ the number of configuration changes in a trajectory, for example the number of spin flips in a chain of spins. Here we show that our method may also be applied if a time-integrated observable is considered. In the following $B(t)$ denotes a static observable that depends only on the configuration of the system at time $t$ and the corresponding time-integrated quantity is
\begin{equation}
\B_{t} = \int^{t}_{0}\text{d}t' B(t').
\end{equation}
The probability of the system being in configuration $\C$ at time $t$, given a certain value of $\B_{t}$, is $P(\C|\B_{t},t)$. We then define
\begin{equation}
P(\C,s,t)=\sum_{\B_{t}}P(\C|\B_{t},t){e}^{-s\B_{t}}
\end{equation}
such that the MGF of $\B_{t}$ may be written
\begin{equation}
Z(s,t)=\sum_{\C}P(\C,s,t)
\end{equation}
In matrix notation, the corresponding vector $\ket{P(s,t)}$ containing the $P(\C,s,t)$'s obeys the modified master equation
\begin{equation}
\label{eq:modmaster}
\frac{d}{dt}\ket{P(s,t)}=\W(s)\ket{P(s,t)},
\end{equation}
with the matrix $\W(s)$ defined as~\cite{Lecomte2007,Garrahan2007,Garrahan2009}
\begin{equation}
\label{eq:Meqs}
\W(s)\equiv \W - s\sum_{\C}b(\C)\ket{\C}\bra{\C}.
\end{equation}
Here $b(\C)$ is the value of the observable $B$ in configuration $\C$ and $\bra{\C}$ is a projection state of configuration $\C$. We proceed by formally solving Eq.~(\ref{eq:modmaster}) as
\begin{equation}
\ket{P(s,t)}={e}^{\W(s)t}\ket{P(0)},
\end{equation}
where $\ket{P(0)}$ is the state of the system at the initial time $t=0$. Introducing the ``flat'' state, $\bra{-}\equiv(1,\ldots,1)$, we can express the MGF as
\begin{equation}
\label{eq:MGF2}
Z(s,t)  = \bra{-}{e}^{\W(s)t}\ket{P(0)}.
\end{equation}
The time evolution of the MGF is determined by the eigenvalues of $\W(s)$. At zero counting field, the matrix $\W(0)$ has a single zero-eigenvalue $\lambda_0(s=0)=0$ corresponding to the stationary state, defined as the normalized solution of $\W(0)\ket{P^{S}}=0$. All other eigenvalues $\lambda_{j\neq0}(s=0)$ have negative real-parts, ensuring exponential relaxation from an arbitrary initial state towards the stationary state. As the counting field is introduced, the eigenvalue spectrum is perturbed, and the long-time limit of the MGF is governed by the eigenvalue with the largest real-part. We then have
\begin{equation}
Z(s,t)\approx{e}^{\max_j[\lambda_j(s)]t},\,\, t\rightarrow\infty
\end{equation}
and the CGF becomes
\begin{equation}
\Theta(s,t)\approx t\max_j[\lambda_j(s)],\,\, t\rightarrow\infty
\end{equation}
from which we can identify
\begin{equation}
\theta(s)=\max_j[\lambda_j(s)]
\end{equation}
as the LD function in Eq.~(\ref{eq:CGFlongtime}). Importantly, the introduction of a counting field makes it possible that two large eigenvalues may cross each other at particular values of the counting field $s=s_c$, where the dynamical free energy becomes singular, signaling a trajectory phase transition. Analogous to equilibrium statistical mechanics, a first-order trajectory phase transition gives rise to a discontinuity in the first derivative of $\theta(s)$ at $s=s_c$. Similarly, if the discontinuity appears at a higher derivative, we may speak of a continuous phase transition.

As we show below, trajectory phase transitions may be inferred from the complex zeros of the MGF at finite times as they move towards the transition value of the counting field $s_c$. In particular, the position of the leading pair of these dynamical Lee-Yang zeros can be extracted from the high-order cumulants of the dynamical observable of interest. This approach was proposed by two of us in Ref.~\cite{Flindt2013}. For completeness, we describe the details of the method in the following section, before turning to a concrete application.

\section{Lee-Yang Zeros Method}
\label{sec:Method}

We consider the MGF close to a transition value of the counting field, where two large eigenvalues are degenerate, $\lambda_0(s_c)=\lambda_1(s_c)$. For $s\simeq s_c$, the MGF in Eq.~(\ref{eq:MGF2}) can be approximated as
\begin{equation}
Z(s,t)  \simeq c_0(s)e^{\lambda_0(s)t}+c_1(s)e^{\lambda_1(s)t},
\end{equation}
where $c_0(s)$ and $c_1(s)$ are given by the initial conditions. The contributions from other eigenvalues are small and can be neglected. Solving for the zeros of the MGF, we find
\begin{equation}\label{eq:BCond}
\lambda_0(s)=\lambda_1(s)+\frac{\log[c_1(s)/c_0(s)]+i\pi(2n+1)}{t},
\end{equation}
with $n$ being an integer. Importantly, this equation reduces to $\lambda_0(s)=\lambda_1(s)$ in the long-time limit, $t\rightarrow\infty$. Thus, with time the dynamical Lee-Yang zeros of the MGF will move towards the transition point at $s=s_c$.

The motion of the Lee-Yang zeroes can be extracted from the high-order cumulants of the dynamical observable of interest. To see this, we express the MGF in terms of the dynamical Lee-Yang zeros. Based on the Hadamard factorization theorem, we expect that the MGF can be written in terms of its zeros as
\begin{equation}
\label{eq:LYFact}
Z(s,t) = e^{a(t)s}\prod_{j}\frac{[s_{j}(t)-s]}{s_{j}(t)}.
\end{equation}
Here $s_{j}(t)$ are the dynamical Lee-Yang zeros and $a(t)$ is a real function of time which is independent of the counting field.  The Lee-Yang zeros come in complex conjugate pairs at all times since the MGF is a real function for real $s$.

Using this expression, we may write the CGF as
\begin{equation}
\Theta(s,t)=a(t)s+\sum_{j}(\log[s_{j}(t)-s]-\log[s_{j}(t)]).
\end{equation}
The cumulants of $\B_t$ are then~\cite{Flindt2009,Flindt2010,Kambly2011}
\begin{equation}
\label{eq:LYCumu}
\llangle \B_t^{n}\rrangle = a(t)\delta_{n,1}+
(-1)^{(n-1)}(n-1)!\sum_{j}\frac{{e}^{-in\arg[{s}_{j}(t)]}}{|s_{j}(t)|^{n}},
\end{equation}
having introduced the polar notation
\begin{equation}
s_j=|s_{j}|e^{i\arg[{s}_{j}]}.
\end{equation}
The zeros of the MGF correspond to logarithmic singularities in the CGF which determine the high-order derivatives of the CGF, or the cumulants, in accordance with Darboux's theorem~\cite{Dingle1973,Berry2005}.

For large $n$, the sum in Eq.~(\ref{eq:LYCumu}) is dominated by the leading pair of Lee-Yang zeros closest to the origin, denoted as $s_{0}(t)$ and $s^{*}_{0}(t)$, and we may approximate the sum as~\cite{Bhalerao2003,Berry2005,Flindt2009,Flindt2010,Kambly2011,Flindt2013}
\begin{equation}
\label{eq:LYApprox}
\llangle\B_t^{n}\rrangle\approx (-1)^{(n-1)}(n-1)!\frac{2\cos[n\arg
s_{0}(t)]}{|s_{0}(t)|^{n}}.
\end{equation}
We see that the higher-order cumulants grow as the factorial of the cumulant order $n$ and oscillate as a function of any parameter that changes the complex argument $\arg s_{0}$ of the dominant pair of Lee-Yang zeros~\cite{Berry2005,Flindt2009}. This behaviour has been observed experimentally in real-time counting experiments on electron transport through a quantum dot~\cite{Flindt2009,Fricke2010b,*Fricke2010a}. Moreover, from the relation above follows the matrix equation~\cite{Zamastil2005,Flindt2010,Kambly2011,Flindt2013}
\begin{equation}
\label{eq:MatEq}
\left[
\begin{array}{cc}
1 & -\frac{\kappa^{(+)}_{n}}{n} \vspace{.05cm} \\
1 & -\frac{\kappa^{(+)}_{n+1}}{n+1}\\
\end{array}
\right]
\cdot
\left[
\begin{array}{c}
-(s_{0}+s^{*}_{0}) \\
|s_{0}|^{2}\\
\end{array}
\right]
=
\left[
\begin{array}{c}
(n-1)\kappa^{(-)}_{n}\vspace{.25cm}\\
n\kappa^{(-)}_{n+1}\\
\end{array}
\right],
\end{equation}
which can easily be solved for the leading pair of Lee-Yang zeros given the ratios of cumulants
\begin{equation}
\kappa^{(\pm)}_{n}(t)\equiv \frac{\llangle\B_t^{n\pm1}\rrangle}{\llangle\B_t^{n}\rrangle}.
\end{equation}
Thus, from a measurement (or simulation) of the high-order cumulants of $\B_t$ at finite times, evolving under the unbiased dynamics at $s=0$, it is possible to monitor the leading pair of Lee-Yang zeros as they move towards a transition value of the counting field at which a trajectory phase transition occurs~\cite{Blythe2002,Bena2005,Flindt2013}.

This method was used by two of us in Ref.~\cite{Flindt2013} to investigate a first-order trajectory phase transition occurring at $s_c=0$ in kinetically constrained models of glass formers~\cite{Fredrickson1984,Jackle1991}. In the following sections, we apply the same method to a system which exhibits a continuous trajectory phase transitions along a whole curve of critical points in the complex plane of the counting field.

\section{Glauber-Ising Chain}
\label{sec:Model}

The one-dimensional Glauber-Ising model consists of a periodic chain of $N$ classical spins with total energy
\begin{equation}
E =-\frac{J}{2}\sum_{i}S_{i}S_{i+1},
\end{equation}
where the sum runs over all sites of the chain and $S_{i} = \pm1$ is the value of the spin on site $i$~\cite{Glauber1963}. The rate for flipping the spin on site $i$ is given as
\begin{equation}
\Gamma_{i}=\frac{\Gamma}{1+e^{\beta \Delta E_i}},
\end{equation}
where
\begin{equation}
\Delta E_i = JS_{i}(S_{i-1}+S_{i+1}),
\end{equation}
is the energy cost of flipping the spin and $\beta=1/k_BT$ is the inverse temperature. The rate $\Gamma$ sets the overall time-scale for the spin-flip processes. The spin-flip rates obey detailed balance, such that the spins become Boltzmann distributed in the stationary state. However, in contrast to its stationary properties, the relaxation of the system towards equilibrium is a complex dynamical process.

In the following, we investigate the dynamical fluctuations of the time-integrated energy, taking the energy function $E(t)$ as the static observable $B(t)$ which forms $\B_{t}$ as the time-integral
\begin{equation}
\E_t=\int^{t}_0\text{d}t' E(t').
\end{equation}
To this end, we evaluate the time-dependent MGF of the time-integrated energy. Technically, we proceed as in Ref.~\cite{Jack2010} by introducing the variables
\begin{equation}
n_{i}=\frac{1}{2}(1-S_{i}S_{i+1})
\end{equation}
corresponding to the number ($n_{i}=0,1$) of domain walls between sites $i$ and $i+1$. In terms of these domain wall variables the energy function simplifies to
\begin{equation}
E = J\sum_{i}(n_{i}-1/2).
\end{equation}
We may moreover express the Master operator $\W$ in terms of Pauli matrices by defining
\begin{equation}
\sigma^z_i=2(n_{i}-1/2)
\end{equation}
together with the standard raising and lowering operators $\sigma^{+}_{i}$ and $\sigma^{-}_{i}$. The presence of a domain wall corresponds to the up-state of $\sigma^z_i$, while no domain wall is represented by the down-state. In this notation, we have
\begin{equation}
E = \frac{J}{2}\sum_{i}\sigma^z_i
\label{eq:energyOp}
\end{equation}
and
\begin{equation}
\Delta E_i = -J(\sigma^z_i+\sigma^z_{i-1}).
\end{equation}
The generator for the time evolution may then be written as (see also Ref.~\cite{Jack2010})
\begin{equation}
\label{eq:WGI}
\W\!
=\!\frac{\Gamma}{2}\sum_{i}\left(2\sigma^{+}_{i}\!\sigma^{-}_{i+1}\!+\!\gamma\sigma^{-}_{i}\!\sigma^{-}_{i+1}\!+\!\lambda\sigma^{+}_{i}\!\sigma^{+}_{i+1}\!+\!(\lambda\!-\!1)\sigma^{z}_{i}\!-\!1\right)\!,
\end{equation}
where
\begin{equation}
\gamma = \frac{2}{1+{e}^{-2J\beta}}
\label{eq:gamma}
\end{equation}
and
\begin{equation}
\lambda = 2-\gamma.
\label{eq:lambda}
\end{equation}
From Eq.~(\ref{eq:Meqs}), the biased Master operator becomes
\begin{equation}
\label{eq:bisaedWGI}
\W(s)=\W-s\frac{J}{2}\sum_{i}\sigma_i^{z}.
\end{equation}
We note that the LD statistics of the time-integrated energy for anti-ferromagnetic interactions ($J\rightarrow -J$) can be obtained by changing the sign of the counting field $s$ and the inverse temperature $\beta$. In the following, energy and time are measured in units of $J$ and $\Gamma^{-1}$, and we are free to set $J=1$ and $\Gamma=1$.

We continue by symmetrizing $\W(s)$ to obtain the non-Hermitian matrix
\begin{equation}\label{eq:QHam}
\Ham(s)={e}^{\beta E/2}\W(s){e}^{-\beta E/2},
\end{equation}
where $E$ is the diagonal energy operator in Eq.~(\ref{eq:energyOp}). The matrix $\Ham(s)$ then takes the form of a non-Hermitian Hamiltonian for a quantum spin chain
\begin{equation}\label{eq:QHam1}
\begin{split}
\Ham(s)=\frac{1}{2}\sum_{i}&\Big(\frac{1+\sqrt{\gamma\lambda}}{2}\sigma^{x}_{i}\sigma^{x}_{i+1}+\frac{1-\sqrt{\gamma\lambda}}{2}\sigma^{y}_{i}\sigma^{y}_{i+1}\\
&+(\lambda-1-s)\sigma^{z}_{i}-1\Big).
\end{split}
\end{equation}
The counting field $s$ is in general complex and appears as a complex magnetic field in the non-Hermitian Hamiltonian above. In particular, one should note that the biased dynamics maps directly to the Hamiltonian of the one-dimensional quantum Ising chain, when $s$ is real and $\beta=0$, such that $\gamma=\lambda=1$.  We now focus on the case, where $s$ is real to evaluate the MGF and CGF in the long-time and large-system-size limit, before analytically continuing to the complex-$s$ plane.

We diagonalize the Hamiltonian by applying a Bogoliubov rotation followed by a Jordan-Wigner transformation (see Appendix~\ref{app:diag}), yielding
\begin{equation}
\label{eq:QHam2}
\Ham(s)=-\sum_{k}\left[\Omega_{k}(s)(c^{\dagger}_{k}c_{k}-1/2)+1/2\right].
\end{equation}
Here $c_k^{\dagger}$ and $c_k$ are fermionic creation and annihilation operators labeled by the wave vectors
\begin{equation}
k=\frac{\pi n}{N},\,\,\,\, n=-N+1,-N+3,\ldots,N-1.
\label{eq:kvalues}
\end{equation}
and the dispersion relation reads
\begin{equation}
\label{eq:disrelat}
\Omega_{k}(s)=\sqrt{(s-\lambda+1-\cos k)^{2} + \gamma\lambda\sin^{2}k}.
\end{equation}
In terms of the Hamiltonian of the quantum spin chain, the MGF becomes
\begin{equation}
Z(s,t) = \bra{0}{e}^{t\Ham(s)}\ket{0},
\end{equation}
where $\ket{0}$ is the ground state of $\Ham(s)$ at $s = 0$.

\begin{figure*}
\includegraphics[width=\textwidth]{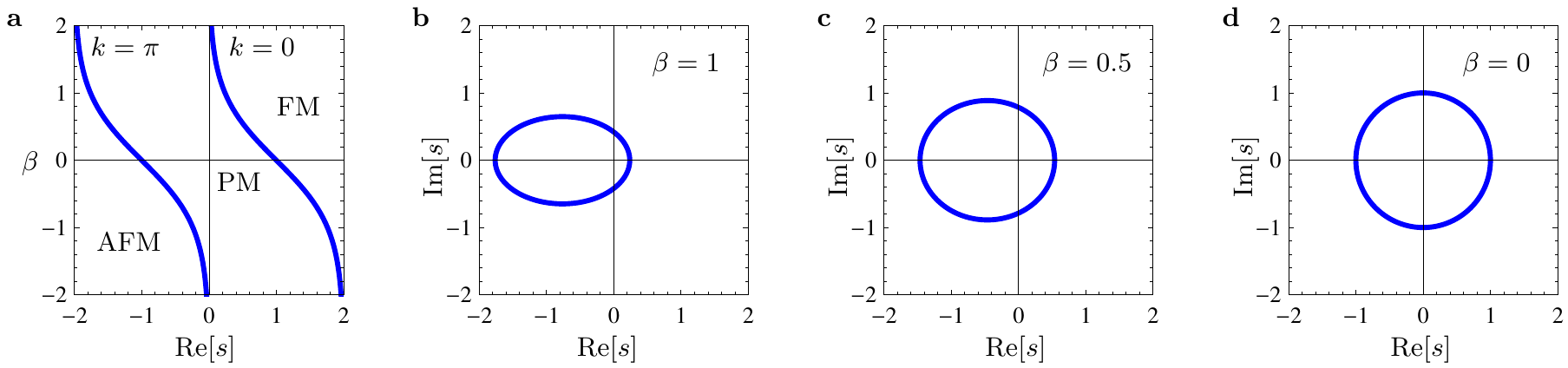}
\caption{(Color online). Trajectory phase diagram and closed curves of critical points. (a) Trajectory phase diagram in the plane of the inverse temperature $\beta$ and the real-part of the counting field $\mathrm{Re}\{s\}$. (Changing the sign of the interaction $J\rightarrow -J$ is equivalent to the sign changes $s\rightarrow -s$ and $\beta\rightarrow -\beta$.) The $k = \pi$ and $k = 0$ modes separate paramagnetically (PM) ordered trajectories from anti-ferromagnetic (AFM) and ferromagnetic (FM) trajectories, respectively. (b-d) Closed curves of trajectory phase transition points in the complex plane of the counting field for different inverse temperatures, $\beta= 1$, 0.5, 0. }
\label{fig:fig1}
\end{figure*}

Using this expression we obtain the MGF as
\begin{equation}
\label{eq:Z}
Z(s,t)=\prod_{k>0}\underbrace{{e}^{t(\Omega_{k}(s)-1)}\cos^{2}\alpha^{s}_{k}(1+\tan^{2}\alpha^{s}_{k}{e}^{-2t\Omega_{k}(s)})}_{Z_{k}(s,t)}
\end{equation}
where the $\alpha^{s}_{k}$'s are related to the angles of the Bogoliubov rotation (see Appendix~\ref{app:diag}), and we have used that $Z_{-k}(s,t)=Z_{k}(s,t)$ . Additionally, the CGF becomes
\begin{equation}
\label{eq:CGF_GI}
\Theta(s,t)=\sum_{k>0}\log Z_{k}(s,t)=\sum_{k>0} \Theta_{k}(s,t),
\end{equation}
showing explicitly that each $k$-mode contributes independently to the fluctuations of the time-integrated energy with a corresponding term $\Theta_{k}(s,t)$ in the CGF.

Equations~(\ref{eq:Z}) and (\ref{eq:CGF_GI}) constitute the central results of this section as they allow us to investigate the time-dependent fluctuations of the time-integrated energy and its cumulants.  As an important check of our result, we see that Eq.~(\ref{eq:Z}) in the long-time limit correctly takes the form $Z(s,t)\approx e^{t\theta(s)}$, where
\begin{equation}
\theta(s)=\sum_{k>0}\underbrace{\left[\Omega_{k}(s)-1\right]}_{\theta_k(s)}=\frac{1}{2}\sum_{k} \theta_k(s)
\label{eq:GImodel_LD}
\end{equation}
is the LD function in agreement with Ref.~\cite{Jack2010}, having corrected for a missing constant of $-1/2$. In the last step, we used $\Omega_{-k}(s)=\Omega_{k}(s)$, allowing us to extend the sum to negative values of $k$. As in Ref.~\cite{Jack2010}, we can take the large-system-size limit, $N\rightarrow \infty$, where the LD function becomes
\begin{equation}
\theta(s)\rightarrow\frac{N}{2}\int_{-\pi}^{\pi}\frac{\text{d} k}{2\pi}(\Omega_{k}(s)-1)
\label{eq:GImodel_thermo}
\end{equation}
from which we may determine the trajectory phase diagram of the Glauber-Ising chain.  In the thermodynamic limit, one should consider the scaled dynamical free energy $\theta(s)/N$, which is well-behaved as $N\rightarrow\infty$.

\section{Trajectory phase diagram}
\label{sec:dpg}

We are now ready to explore the trajectory phase diagram of the Glauber-Ising model based on the LD function found above. The following analysis simplifies considerably if we consider the individual $k$-mode contributions $\theta_k(s)$ to the LD function, see Eq.~(\ref{eq:GImodel_LD}). We first treat the counting field $s$ as a real variable. The individual $k$-mode contributions $\theta_k(s)$ are singular at values of the counting field, where $\Omega_{k}(s)$ has a square-root branch point. These can be found by solving $\Omega_{k}(s)=0$, which immediately gives us a series of phase transition points that we denote as $s_c$. Requiring that the counting field $s$ is real, we find $k=0,\pi$ and
\begin{equation}\label{eq:PTCurve}
s_c^{(\pm)}(\beta)=\lambda-1\pm 1,
\end{equation}
where $\lambda$ is given by Eqs.~(\ref{eq:gamma}) and~(\ref{eq:lambda}). The $k = 0$ and $k=\pi$ modes separate paramagnetically ordered trajectories from anti-ferromagnetic and ferromagnetic trajectories, respectively \cite{Jack2010}. The two lines of trajectory phase transition points are shown in the left panel of Fig.~\ref{fig:fig1} together with the different trajectory phases. We note that the phase transitions are continuous~\cite{Jack2010}.

Next, we promote the counting field $s$ to a complex variable and solve again for the trajectory phase transition points of the individual $k$-mode contributions. In this case, all $k$-mode contributions are singular at some complex value of the counting field. We now find that $\Omega_{k}(s)$ has square-root branch points at the following complex values of the counting field:
\begin{equation}\label{eq:PTCurve}
s_c(k,\beta)=\lambda-1+\cos{k}+i\sqrt{\gamma\lambda}\sin{k}.
\end{equation}
In the thermodynamic limit, $N\rightarrow\infty$, the wave vector $k\in [-\pi,\pi]$ becomes continuous such that the transition points (for a fixed temperature) form a closed curve in the complex plane of the counting field, see Fig.~\ref{fig:fig1}. At finite temperatures ($\beta>0$), the curve is an ellipse, since $\gamma\lambda\neq1$. However, as the temperature increases, the curves approach the unit circle as expected in the infinite-temperature limit ($\beta=0$ and $\gamma=\lambda=1$) according to the Lee-Yang circle theorem of the associated one-dimensional quantum Ising model~\cite{Yang1952}.

\begin{figure*}
\includegraphics[width=0.85\textwidth]{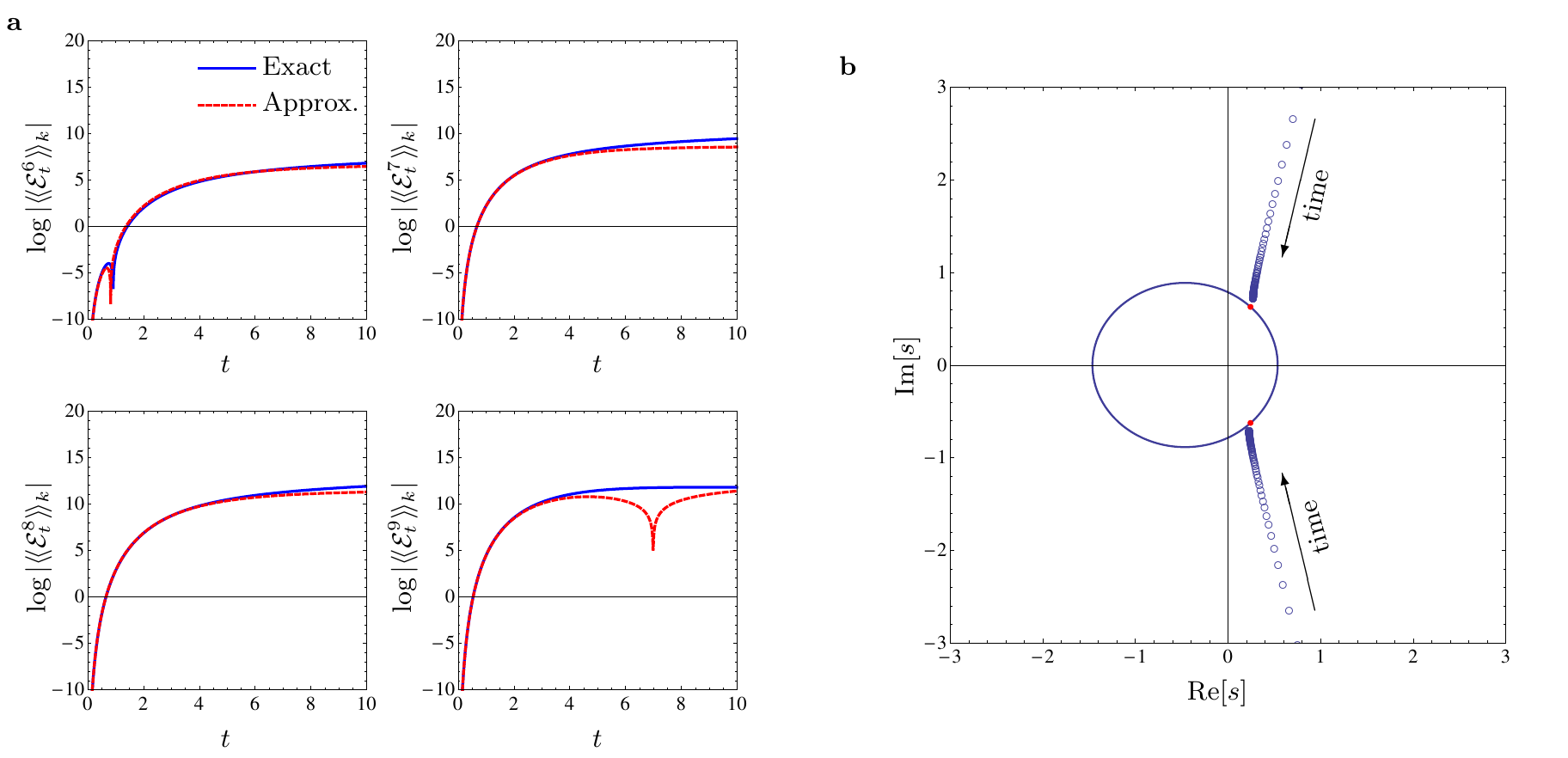}
\caption{(Color online). Mode-resolved cumulants and dynamical Lee-Yang zeros. (a) Mode-resolved cumulants of order $n=6$-$9$ as functions of time with $k=3\pi/4$ and $\beta=0.5$. Exact results (full lines) together with the approximation (dashed lines) based on the extracted pair of Lee-Yang zeros. (b) Motion of the Lee-Yang zeros (open circles) extracted from the high-order cumulants. The Lee-Yang zeros move towards the phase transitions points (red circles) on the closed curve of transition points.}
\label{fig:fig2}
\end{figure*}

\section{Mode-resolved cumulants}
\label{sec:kres}

Having understood the trajectory phase diagram, we move on to the application of the method proposed in Ref.~\cite{Flindt2013} and described again in Sec.~\ref{sec:Method}. We recall that the method is applicable in situations, where only the real physical dynamics (taking place without the counting field, i.~e.~$s=0$) can be probed during a finite period of time. The purpose of this section is to apply the proposed method in order to detect signatures of the trajectory phase transition points found above.

We first consider the individual $k$-mode contributions $\Theta_k(s,t)$ at finite times, see Eqs.~(\ref{eq:Z}) and (\ref{eq:CGF_GI}). The corresponding (time-dependent) $k$-resolved cumulants are
\begin{equation}
\label{eq:k_cumu}
\llangle \E_t^{n}\rrangle_k = (-1)^{n}\partial^{n}_{s}\Theta_k(s,t)|_{s\rightarrow0}.
\end{equation}
In the left panel of Fig.~\ref{fig:fig2}, we show the $k$-resolved cumulants (full lines) of order $m=6,7,8,9$ as functions of time for a particular value of $k$. (We have checked that other values of $k$ give similar results.) The absolute value of the cumulants is plotted on a logarithmic scale, such that a downwards-pointing spike corresponds to a cumulant crossing zero.

In the next step, we use Eq.~(\ref{eq:MatEq}) to extract the leading pair of Lee-Yang zeros closest to $s=0$ from the finite-time cumulants. This allows us to follow the motion of the leading pair of Lee-Yang zeros as they move with time. In the right panel of Fig.~\ref{fig:fig2}, we show the motion of the extracted Lee-Yang zeros (open circles) in the complex plane of the counting field. Together with the Lee-Yang zeros, we plot the full curve of critical points (full line) as well as the particular critical points corresponding to the given value of $k$ (red circles). Clearly, the leading pair of Lee-Yang zeros approaches the critical points with increasing time. In particular, we see that the non-zero critical points can be deduced from the finite-time behaviour of the high-order cumulants (obtained at zero counting field).

To corroborate our extraction of the leading pair of Lee-Yang zeros, we plug the zeros back into Eq.~(\ref{eq:LYApprox}) and show the resulting curves (dashed lines) in the left panel of Fig.~\ref{fig:fig2} together with the exact results (full lines). The agreement between the two sets of curves demonstrates that Eq.~(\ref{eq:LYApprox}) provides a good approximation to the exact results. In general, we expect that Eq.~(\ref{eq:LYApprox}) works well at relatively short times, where only a single pair of  Lee-Yang zeros is close to $s=0$. Some deviations are already seen in Fig.~\ref{fig:fig2} as the second pair of Lee--Yang zeros comes close to $s=0$ and start to contribute to the sum in Eq.~(\ref{eq:LYCumu}). The accuracy of the method may be improved by using higher cumulants \cite{Zamastil2005,Flindt2010,Kambly2011}. At longer times, more Lee-Yang zeros move towards the critical points and the agreement breaks down between Eq.~(\ref{eq:LYApprox}) and the exact results for the high-order cumulants (not shown). However, as seen in the right panel of Fig.~\ref{fig:fig2}, the critical points can be extracted from the $k$-resolved cumulants before this breakdown.

\section{Full analysis}
\label{sec:MGFanaly}

\begin{figure*}
\includegraphics[width=0.8\textwidth]{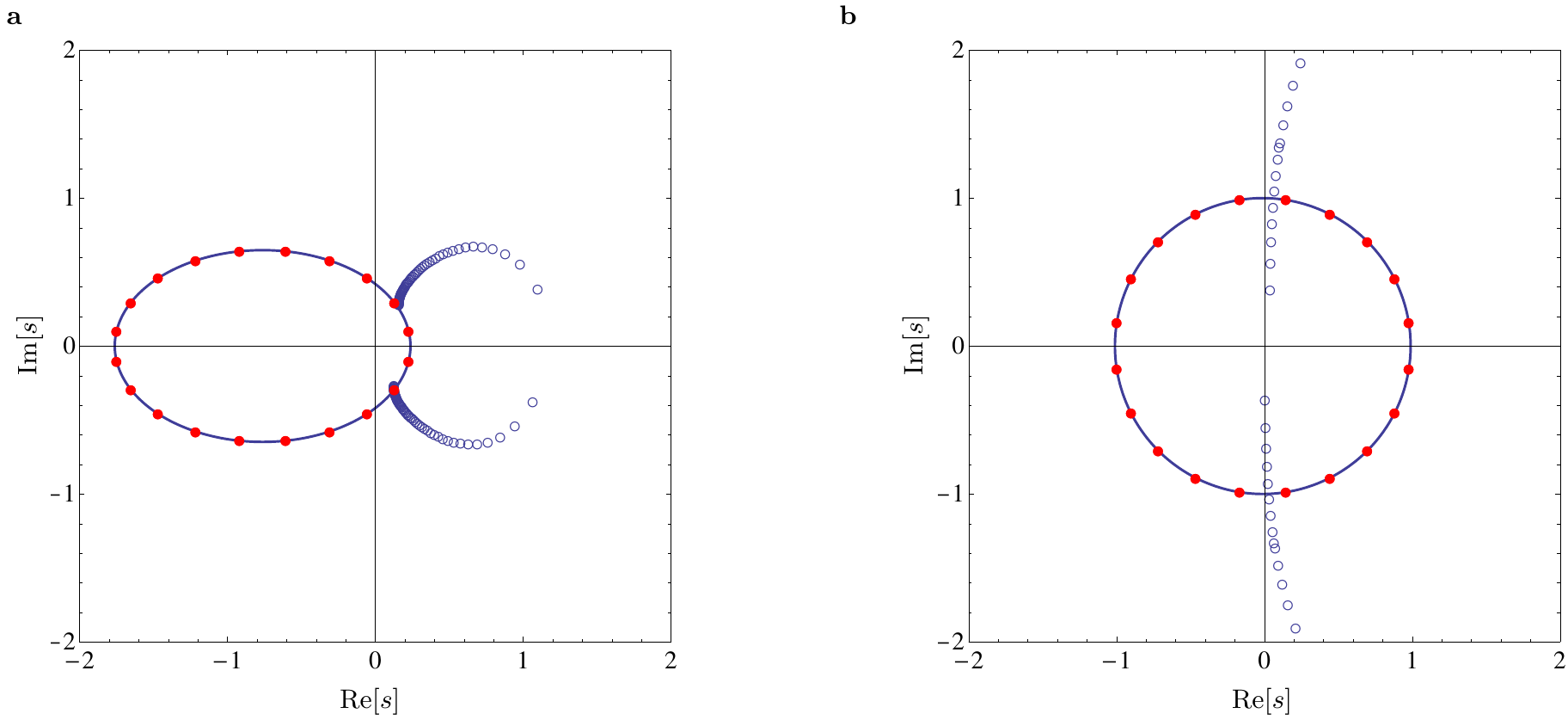}
\caption{(Color online). Dynamical Lee-Yang zeros at different temperatures for a chain with $N=20$. (a) At low temperatures ($\beta=1$), the fluctuations of the time-integrated energy are dominated by the long-wavelength modes, whose singularities are closest to $s=0$. The extracted pair of Lee-Yang zeros (open circles) move with time towards these points.  (b) In the high-temperature limit ($\beta=0$), all modes contribute to the fluctuations and the corresponding singularities are equally distant from $s=0$. In this case, the extracted pair of Lee-Yang zeros (open circles) do not move towards the phase transition points. The dynamical Lee-Yang zeros were extracted from the cumulants of order $n=6,7,8,9$.}
\label{fig:fig3}
\end{figure*}

We now consider the full MGF in Eq.~(\ref{eq:Z}) and the corresponding cumulants which are simply the sum of all $k$-resolved cumulants
\begin{equation}
\label{eq:full_cumu}
\llangle \E_t^{n}\rrangle = \sum_k \llangle \E_t^{n}\rrangle_k.
\end{equation}
We recall that the approximation in Eq.~(\ref{eq:LYApprox}) only includes the leading pair of Lee-Yang zeros, which will converge to (at most) two distinct points in the complex-$s$ plane. This immediately makes it is clear that one cannot determine the full line of phase transition points as with the $k$-resolved cumulants.  However, in some cases a few critical points closest to $s = 0$ may dominate the fluctuations of the time-integrated energy.

Indeed, considering again Fig.~\ref{fig:fig1} it is clear that the fluctuations of the time-integrated energy in the high-temperature limit ($\beta=0$) are influenced by all the critical points on the unit circle which are equally far from $s=0$. In contrast, as the temperature is lowered, a few points on the ellipse are closest to the origin and those points will dominate the dynamics of the system and the fluctuations of the time-integrated energy. With this in mind, we may anticipate that our method will work well in the low-temperature regime. This regime has interesting dynamical properties as thermal fluctuations are suppressed and only the intrinsic dynamical fluctuations associated with the model determine the temporal evolution of the cumulants. In contrast, we expect that the method will not be successful in capturing the critical points in the high-temperature limit.

In Fig.~\ref{fig:fig3}, we show the dynamical Lee-Yang zeros extracted from the high-order cumulants of the time-integrated energy at low and at high temperatures, respectively. For these calculations, we have used the full cumulants and not the $k$-resolved cumulants as in the previous section. At low temperatures (left panel), a few critical points on the ellipse are closest to the origin and these points are expected to dominate the fluctuations of the time-integrated energy. In this case, we see that our method is successful in capturing the motion of the leading pair of Lee-Yang zeros as they with time approach the dominating critical points. At low temperatures, the high-order cumulants are dominated by the low-$k$ modes, whose singularities are closest to $s=0$. Going to the high-temperature limit ($\beta=0$) in the right panel of Fig.~\ref{fig:fig3}, this picture changes drastically. In this case, all $k$-modes contribute significantly to the fluctuations and all critical points become important. Of course, we can in principle still extract dynamical Lee-Yang zeros from the high-order cumulants using Eq.~(\ref{eq:MatEq}). However, as Eq.~(\ref{eq:LYApprox}) is no longer valid, since several Lee-Yang zeros now contribute to the high-order cumulants, the extraction is no longer meaningful. Indeed, as we see in the right panel of Fig.~\ref{fig:fig3}, the extracted Lee-Yang zeros do not approach the critical points as time increases.

\section{Conclusions}
\label{sec:Conc}

We have employed our recently proposed method to extract dynamical Lee-Yang zeros from high-order cumulants of dynamical observables. Concretely, we have investigated the  time-integrated energy of the one-dimensional Glauber-Ising model for which we evaluated the generating function at finite times by mapping the generator of the biased dynamics to a non-Hermitian Hamiltonian of an associated quantum spin chain. The Glauber-Ising chain is of special interest, since the singularities of the dynamical free energy make up a whole curve of critical points and the trajectory phase transitions are continuous rather than of first order. Compared to our previous work on dynamical systems with a single first-order transition point at zero counting field~\cite{Flindt2013}, this makes the Glauber-Ising model a particularly challenging problem for the proposed method.

As we have shown, the full trajectory phase space diagram may still be reconstructed using mode-resolved cumulants. Considering the full cumulants, we find that the dominating singularities of the system can still be extracted, but only at low temperatures, where the fluctuations are dominated by the long-wavelength modes. In contrast, at high temperatures all modes are important, making it difficult to extract the transition points from the cumulants.

Our work leaves a number of interesting questions and tasks for future research. It would be interesting to see if the nature of a trajectory phase transition (first-order or continuous) can be understood from the dynamical Lee-Yang zeros. In analogy with conventional Lee-Yang theory,  we expect that the way the dynamical Lee-Yang zeros accumulate in the long-time limit carries this information~\cite{Blythe2002}. Such an approach may require that not only the leading pair of Lee-Yang zeros is extracted from the high-order cumulants, but also some of the sub-leading pairs may be needed. Finally, it would be interesting to apply the proposed method to a number of open quantum systems, for instance driven quantum dots~\cite{Garrahan2010}, micromasers~\cite{Garrahan2011}, and single-electron transistors~\cite{Genway2012}.

\section{acknowledgements}
This work was supported by Swiss NSF as well as by EPSRC Grant no.~EP/I017828/1 and Leverhulme Trust grant no.~F/00114/BG.

\appendix

\section{Diagonalisation of $\Ham_{s}$}
\label{app:diag}
The quantum spin chain is governed by the Hamiltonian $\Ham(s)$ in Eq.~(\ref{eq:QHam1}), which can be diagonalized  using a Jordan-Wigner transformation in combination with a Bogoliubov rotation~\cite{Sachdev2011}.  The Jordan-Wigner transformation expresses the spin-$\nicefrac{1}{2}$ operators $\sigma^{z}_{i}$, $\sigma^{+}_{i}$, and $\sigma^{-}_{i}$  at site $i$ in terms of corresponding fermionic operators ${a}_{i}$ and ${a}^{\dagger}_{i}$ with $\{{a}^{\dagger}_{i},{a}_{j} \}=\delta_{i,j}$ as
\begin{equation}
\label{eq:JWT}
\begin{split}
\sigma^{z}_{i} &= 1-2a^{\dagger}_{i}a_{i},\\
\sigma^{+}_{i} &= \prod_{j<i}(1-2a^{\dagger}_{j}a_{j}) a_{i}, \\
\sigma^{-}_{i} &= \prod_{j<i}(1-2a^{\dagger}_{j}a_{j}) a^{\dagger}_{i}.
\end{split}
\end{equation}
Moreover, by changing to the Fourier representation
\begin{equation}
a_{i} = \frac{1}{\sqrt{N}}\sum_{k}{e}^{-ikr_{i}}a_{k}
\end{equation}
we may rewrite the Hamiltonian as
\begin{equation}
\label{eq:InterHam}
\begin{split}
\Ham(s) = \frac{1}{2}\sum_{k}&\Bigl(2(s+1-\lambda)a^{\dagger}_{k}a_{k} + 2\cos k a^{\dagger}_{k}a_{k}\Bigr. \\
&\Bigl.-i\sqrt{\gamma\lambda}\sin k (a_{-k}a_{k}+a^{\dagger}_{-k}a^{\dagger}_{k}-(s+2-\lambda)\Bigr).
\end{split}
\end{equation}
For an even number of spins $N$ and with periodic boundary conditions, the discrete wave vector $k$ takes on the values given by Eq.~(\ref{eq:kvalues}).

We note that the Hamiltonian in Eq.~(\ref{eq:InterHam}) contains terms that do not conserve the number of fermions, for instance $a^{\dagger}_{-k}a^{\dagger}_{k}$. These terms are eliminated next via a canonical Bogoliubov rotation~\cite{Sachdev2011}. This transformation expresses the Jordan-Wigner operators as a linear combination of a new set of $s$-dependent fermionic operators $c_{k,s}$ and $c^{\dagger}_{k,s}$ with $\{c_{k,s},c^{\dagger}_{k',s}\}=\delta_{k,k'}$ as
\begin{equation}\label{eq:BogRot}
\begin{split}
a_{k} &= \cos \phi^{s}_{k} c_{k,s} + i\sin \phi^{s}_{k} c^{\dagger}_{-k,s},\\
a^{\dagger}_{k} &= \cos \phi^{s}_{k} c^{\dagger}_{k,s} - i\sin \phi^{s}_{k}c_{-k,s}.
\end{split}
\end{equation}
The Bogoliubov angles $\phi^{s}_{k}$ satisfy
\begin{equation}
\phi^{s}_{-k}=-\phi^{s}_{k}
\end{equation}
and are chosen such that only terms that conserve the number of fermions are present in the transformed Hamiltonian. To enforce this condition, the Bogoliubov angles must satisfy
\begin{equation}
\tan{\phi_{k}^{s}} =-\frac{\sqrt{\gamma\lambda}\sin{k}}{s+1-\lambda-\cos{k}}.
\end{equation}
With this choice we arrive at Eq.~(\ref{eq:QHam2}) (having omitted the $s$-dependence of the operators $c_{k,s}$ and $c^{\dagger}_{k,s}$) with the free-fermion dispersion relation given by Eq.~(\ref{eq:disrelat}).

Next, we evaluate the MGF $Z(s,t)$. To this end, we note that the vacuum state $\ket{0}_{s=0}$ of $\Ham(s=0)$ may be expressed as a BCS state of $\Ham(s)$ with $s\neq0$.  Since $c_{k,s=0}\ket{0}_{s=0} = 0$ for all $k$, we may expand $\ket{0}$ as a BCS state of the $s$-vacuum $\ket{0}_{s}$ of the form
\begin{equation}
\begin{split}\label{eq:BCS}
\ket{0}_{s=0} &= \frac{1}{\mathcal{N}} \exp\left(\sum_{k>0}
B(k)c^{\dagger}_{k,s}c^{\dagger}_{-k,s}\right)\ket{0}_{s} \\
 &=\bigotimes_{k>0} \left[ \cos{{\alpha}_{k}^{s}}{\ket{0_{k},0_{-k}}}_{s} -i\sin{{\alpha}_{k}^{s}}{\ket{1_k,1_{-k}}}_{s} \right],
\end{split}
\end{equation}
where the second line is obtained by expanding the exponential and using the Bogoliubov rotation~\ref{eq:BogRot}. Here $\mathcal{N}$ is a normalization factor, $\bigotimes$ is a direct product, ${\ket{n_{k},n_{-k}}}_{s}$ indicates occupation states of the fermionic modes with $\pm k$, and the coefficients $\alpha_{k}^{s}$ are related to the Bogoliubov angles via
\begin{equation}
\alpha_{k}^{s} = \frac{{\phi}^{0}_{k}-{\phi}^{s}_{k}}{2}.
\end{equation}
With these expressions at hand, we may evaluate the MGF directly and thereby arrive at Eq.~(\ref{eq:Z}).

\bibliography{GI}
\end{document}